# Interference enhancement of Raman signal of graphene

Y. Y. Wang, Z. H. Ni, and Z. X. Shen*

*Division of Physics and Applied Physics, School of Physical & Mathematical Sciences, Nanyang Technological University, Singapore 637371*

H. M. Wang, Y. H. Wu

*Department of Electrical and Computer Engineering, National University of Singapore, 4 Engineering Drive 3, Singapore 117576*

**Abstract**

Raman spectroscopic studies of graphene have attracted much interest. The G-band Raman intensity of a single layer graphene on Si substrate with 300 nm $SiO_2$ capping layer is surprisingly strong and is comparable to that of bulk graphite. To explain this Raman intensity anomaly, we show that in addition to the interference due to multiple reflection of the incident laser, the multiple reflection of the Raman signal inside the graphene layer must be also accounted for. Further studies of the role of $SiO_2$ layer in the enhancement Raman signal of graphene are carried out and an enhancement factor of ~30 is achievable, which is very significant for the Raman studies. Finally, we discuss the potential application of this enhancement effect on other ultra-thin films and nanoflakes and a general selection criterion of capping layer and substrate is given.

*Corresponding author email: zexiang@ntu.edu.sg

Graphene has attracted a lot of interest since discovered in 2004.[1] The high crystal quality,[2] ballistic transport under ambient condition and the massless Dirac fermions-like charge carriers [3,4] make graphene promising candidate both for future electronics and understanding of fundamental physics.

Raman spectroscopy is a fast, non-destructive way to probe structural and electronic characteristics of graphite materials.[5] Raman spectroscopic studies of graphene have revealed very interesting phenomena, i.e. the Raman second order (2D) band can be used as a simple and efficient way to determine the single layer graphene,[6,7] the Raman spectra are also used to measure the electron/hole dopants in graphene,[8,9] even the electronic structure of bilayer graphene is probed by resonant Raman scattering.[10]

Graphene samples currently used in experiments are usually fabricated by micromechanical cleavage of graphite and transferred to certain substrates which make it visible, typically the Si wafer with 300 nm $SiO_2$ capping layer ($SiO_2$/Si substrate).[11-15] The very strong Raman signal from graphene (single and few layers) on $SiO_2$/Si substrate has been observed by many researchers and this makes the Raman studies of graphene available. [6,7,12] It has been observed by D.Graf et al.[7] and A.Gupta et al.[12] that the G-band Raman intensity would increase almost linearly with the number of graphene layers. In this manuscript, we report on Raman intensity of G-band as a function of number of layers on 300 nm $SiO_2$/Si substrate. It is found that the G-band Raman signal of graphene sheets would decrease when the layers of graphene exceed certain value (~10 layers), while the signal of single layer graphene is comparable to that of graphite, in other words, the Raman signal of single layer graphene is significantly enhanced. We believe this Raman intensity anomaly is because of the interference of laser in graphene sheets as well as the multi-reflection of Raman light. Model based on above opinion is set up and the role of $SiO_2$ capping layer for this enhancement is further investigated. It is found that the maximum enhancement factor (the ratio of Raman signal of single layer graphene with $SiO_2$ capping layer to without) could be ~30, which is significant for the Raman studies.

The graphene samples are prepared by micromechanical cleavage and transferred to Si wafer with ~300 nm $SiO_2$ capping layer.[1] Optical microscopy is used to determine the thickness of graphene sheets and further confirmed by Raman spectra/image and contrast spectra. Details of reflection and contrast experiment are the same as described in reference 13. The Raman spectra are carried out with a WITEC CRM200 Raman system with 532 nm (2.33 eV) excitation and laser power at sample below 0.1 mW to avoid laser induced heating. All the Raman spectra of graphene/graphite sheets are recorded under the same conditions (integration time, laser power, focus, et al.), and each Raman spectrum is an average from several spots on the same sample. A 100× objective lens with a NA=0.95 is used both in the Raman and reflection experiments.

Figure 1(a) gives typical Raman spectra of graphene and graphite. The major Raman features of graphene and graphite are the so called G-band (~1580 cm$^{-1}$) and 2D-band (~2680 cm$^{-1}$), where the 2D-band originates from a double resonance Raman process.[5] The G-band originates from in-plane vibration of sp$^2$ carbon atoms is a doubly degenerate (iTO and LO) phonon mode ($E_{2g}$ symmetry) at the Brillouin zone center.[16] Therefore, the Raman intensity of G-band should be proportional to the thickness of graphene/graphite samples up to the laser penetration depth, which is around 50 nm (~150 layers) for 532 nm laser. Figure 1(b) shows the G-band Raman intensity of graphene sheets as a function of number of layers. The thickness of graphene sheets are determined by contrast spectrum.[13-15] Besides the graphene sheets with one, two, three, five, six, eight, and nine layers, samples *a- h* are more than 10 layers[13-15] whose thickness are ~5 nm and ~50 nm, respectively, as determined by atomic force microscope (AFM). Interestingly, it can be seen that the Raman intensity increases almost linearly until ~10 layers (~ 3nm) and decreases for thicker graphite sheets. Besides, it is obvious that Raman signal of bulk graphite is weaker than that of bilayer graphene. These results deviate from the above understanding that laser penetration depths decide the thickness where highest Raman intensity can be obtained. Therefore, some mechanism must be responsible for this phenomenon.

Need to say that multi-reflection/interference effect on contrast has been well documented but the multiple reflection of Raman is not.[13-15] By considering the multilayer interference of incident light as well as the multi-reflection of Raman signals in graphene/graphite based on the Fresnel's equations,[13,14,15,17] the above Raman results can be explained. Consider the incident light from air ($n_0$=1) onto a graphene sheet, $SiO_2$/ Si trilayer system, as shown in Fig. 2(a). $\tilde{n}_1$=2.6-1.3i, $n_2$=1.46, $\tilde{n}_3$=4.15-0.044i, are refractive indices of graphite, $SiO_2$, and Si at 532 nm, respectively. $d_1$ is the thickness of graphene which is estimated as $d_1=N\Delta d$, and $\Delta d$ =0.335nm,[18,19] where $N$ is the number of layers. $d_2$ is the thickness of $SiO_2$ and the Si substrate is considered as semi-infinite. When a beam of light front at an interface, for example, the air/graphene or graphene/$SiO_2$ interface, a portion of the beam is reflected and the rest is transmitted, thus, an infinite number of optical paths are possible. The G-band Raman intensity of graphene sheets depends on the electric field distribution, which is a result of interference between all these transmitted optical paths in graphene/graphite sheets.

The total amplitude of the electric field at certain depth $y$ in graphene/graphite sheets is viewed as a sum of the infinite transimitted laser, as schematically shown in Fig 2(a), whose amplitudes after considering absorption are:

$$t_1 e^{\beta y} \cdot e^{-i\frac{2\pi \tilde{n}_1 y}{\lambda}},$$

$$t_1 \cdot r' \cdot e^{\beta \cdot (2d_1-y)} \cdot e^{-i\cdot\frac{2\pi \tilde{n}_1 \cdot (2d_1-y)}{\lambda}},$$

$$-t_1 e^{\beta y} \cdot e^{-i\frac{2\pi \tilde{n}_1 y}{\lambda}} \cdot r_1 r' e^{-2i\cdot f_{i1}} \cdot e^{2\beta \cdot d_1},$$

$$-t_1 \cdot r' \cdot e^{\beta \cdot (2d_1-y)} \cdot e^{-i\cdot\frac{2\pi \tilde{n}_1 \cdot (2d_1-y)}{\lambda}} \cdot r_1 r' e^{-2i\cdot f_{i1}} e^{2\beta d_1}, \ldots \quad (1)$$

where $\beta = \frac{-2\pi \cdot k_1}{\lambda}$, ($k_1$ =1.3 is extinction coefficient of graphite and $\lambda$ is the excitation wavelength) emerges as a measure of the absorption in the graphene layers, $t_1 = \frac{2n_0}{n_0+\tilde{n}_1}$ is transmission coefficients at the interface of air/graphene, $r_1 = \frac{n_0-\tilde{n}_1}{n_0+\tilde{n}_1}$ is reflection coefficient at the interface of air/graphene, $fi_{1,2} = \frac{2\pi \cdot \tilde{n}_{1,2} \cdot d_{1,2}}{\lambda}$, are the phase differences when light passes through graphene layers and SiO$_2$ layers, respectively. Here, $r' = \frac{r_2 + r_3 \cdot e^{-2i \cdot fi_2}}{1 + r_2 \cdot r_3 \cdot e^{-2i \cdot fi_2}}$ is the effective reflection coefficient of graphene/(SiO$_2$ on Si) interface, where $r_2 = \frac{\tilde{n}_1 - n_2}{\tilde{n}_1 + n_2}$, $r_3 = \frac{\tilde{n}_2 - \tilde{n}_3}{\tilde{n}_2 - \tilde{n}_3}$ are individually reflection coefficient at the interface of graphene/SiO$_2$ and SiO$_2$/Si.

Thus, the total amplitude of the electric field at the depth $y$ is

$$t = \frac{t_1 \cdot e^{\beta \cdot y} \cdot e^{-i \cdot \frac{2\pi \tilde{n}_1 \cdot y}{\lambda}} + t_1 \cdot r' \cdot e^{\beta \cdot (2d_1 - y)} \cdot e^{-i \cdot \frac{2\pi \tilde{n}_1 \cdot (2d_1 - y)}{\lambda}}}{1 + r_1 r' \cdot e^{-2i \cdot fi_1} \cdot e^{2 \cdot \beta \cdot d_1}} \quad (2)$$

In addition, further consideration should be applied to the multi-reflection of scattering Raman light in graphene/graphite at the interface of graphene/air and graphene/(SiO$_2$ on Si), which contribute to the detected Raman signal. Fig 2(b) schematically shows multi-reflection of the scattering light in certain depth $y$ in graphene sheets. Thus, the detected signal is a result of summation of infinite transmitted light from the interface of graphene/air, which makes the amplitude multiplied by

$$\gamma = \frac{(e^{\beta y} + r' e^{\beta(2d_1 - y)}) t_1'}{1 + r_1 r' e^{2\beta d_1}} \quad (3)$$

where $t_1' = \frac{1 - r_1^2}{t_1}$ represent the transmission coefficients at the interface of

graphene/ air. In the equation above interference of Raman scattering light is not considered as the phase of spontaneous Raman lights are random.

Thus, the total Raman signal as result of interference of infinite transmitted laser optical paths in graphene film followed by considering multi-reflection of Raman scattering light can be expressed as:

$$I = \int_0^{d_1} |t \cdot \gamma|^2 \Delta y \tag{4}$$

Fig. 2(c) shows the calculation results of Raman intensity of G band as a function of number of graphene/graphite layers. The black curve shows the calculation result without considering the multi-reflection of Raman light in graphene, which is given by $I = \int_0^{d_1} |t|^2 \Delta y$. It can be seen that the Raman scattering intensity is strongest at ~38 layers and the intensity of bulk graphite (~1000 layers) is much stronger than that of single and bilayer graphene which is not agreed with experimental results. The red curve gives the calculated results based on equation (4) which is after considering the multi-reflection of Raman light in graphene. In that case, the intensity is strongest at ~ 22 layers and the Raman signal of bulk graphite (~1000 layers) is weaker than that of bilayer graphene, which agrees very well with the experiment data. Thus, the consideration of the multi-reflection of scattering Raman light in graphene/graphite is necessary. The deviation of thickness of highest intensity (~22 layers) from the experimental results (~10 layers) may because, in our simulation results, we don't integrate all possible incident angles and polarizations as well as a weight distribution accounting for the Gaussian beam profile used in experimental. Our calculation for Raman intensity of graphene can also be applied to other thin films and materials.

Attention is then paid to the SiO$_2$ capping layer because of it is believed to be the reason for enhancement of Raman signal of single layer graphene. Fig. 3 shows the enhancement factor E as a function of $\frac{n_{SiO_2} \cdot d_2}{\lambda}$. E is a ratio of Raman intensity of single layer graphene on Si substrate with SiO$_2$ capping layer to without, which is calculated based on equation (4). It can be seen that, E value oscillates periodically with the thickness of SiO$_2$ and the maximum enhancement happens at $\frac{n_{SiO_2} \cdot d_2}{\lambda} \approx 1/4$, 3/4 and so on. The highest enhancement value, for example, when $\frac{n_{SiO_2} \cdot d_2}{\lambda} \approx 3/4$, (d$_2$=269 nm), can be ~30. In our experiments, for SiO$_2$ of 300 nm, E is ~16, which contributes to the high intensity of single layer graphene compared to bulk graphite. In order to verify this enhancement effect, we have also measured the G-band signal of single layer graphene on quartz and SiC substrate, and their signals are <10% of that on SiO$_2$/Si substrate.

The idea of using interference to enhance Raman scattering signal was proposed by R. J. Nemanich et al. and applied on thin absorbing films.[20,21] Enhancement of Raman signal was achieved by efficient use of incident beam and scattering radiation realized by using a substrate consisting a mirror coated with a thin dielectric film of appropriate thickness. In this paper, we deduce the theoretical equation for calculating the scattered Raman intensity as a function of graphene thickness. The electric field distribution in graphene as well as the multi-reflection of scattered Raman signal are taking into account. Our results explain the strong Raman signal of single layer graphene as well as the thickness dependence of Raman intensity. The theoretical calculation matches the experimental results very well.

From above calculation results, it can be seen that a general selection of capping layer and substrate for enhancement can be achieved by choosing $n_1 \gg n_2 \ll n_3$, plus $n_2 d_2 = \frac{\lambda}{4}, \frac{3\lambda}{4}$, and so on.( $n_1$, $n_2$, $n_3$ are individually refractive indices of sample, capping layer and substrate). A total phase change of $2\pi$ ($\pi$ at the interface of capping layer/ substrate plus $\pi$ due to the double thickness of capping layer) can be achieved by this kind of configuration. Therefore, infinite transmitted light in graphene layers have no phase change will enhance Raman signal greatly. Another equivalent choice for achieving this enhancement is $n_1 \ll n_2 \gg n_3$ with $n_2 d_2 = \frac{\lambda}{4}, \frac{3\lambda}{4}$ and so on. . However, choice of refractive indices, $n_1 \gg n_2$, $n_2 \ll n_3$ is more achievable. Besides $SiO_2$, Other dielectrics with lower refractive index such as PMMA (n=~1.49), $CaF_2$ (n=~1.437), LiF(n=~1.39), KBr(n=~1.571),[22] also can be used as capping layer and semiconductors such as Ge(n=~4.24), GaSb(n=~4.31), InAs(n=~4.17) with higher refractive index can be used as substrate.[22]

In summary, we have carried out the Raman measurements on graphene of different thickness on 300 nm $SiO_2$/Si substrate. The anomaly in G-band Raman intensity of graphene sheets is explained by considering the multiple reflection of Raman signal inside the graphene layer as well as the interference effect due to the multiple reflection of the incident laser. Raman signal of single graphene sheet can be enhanced ~30 times by using a 269 nm $SiO_2$ above Si substrate. This enhancement effect can be also applied to other ultrathin sheet-like samples, providing a general method to enhance the Raman signal from flat sheet samples or nanoflakes.

The authors acknowledge Dr. Liu Yanhong for technical assistance.

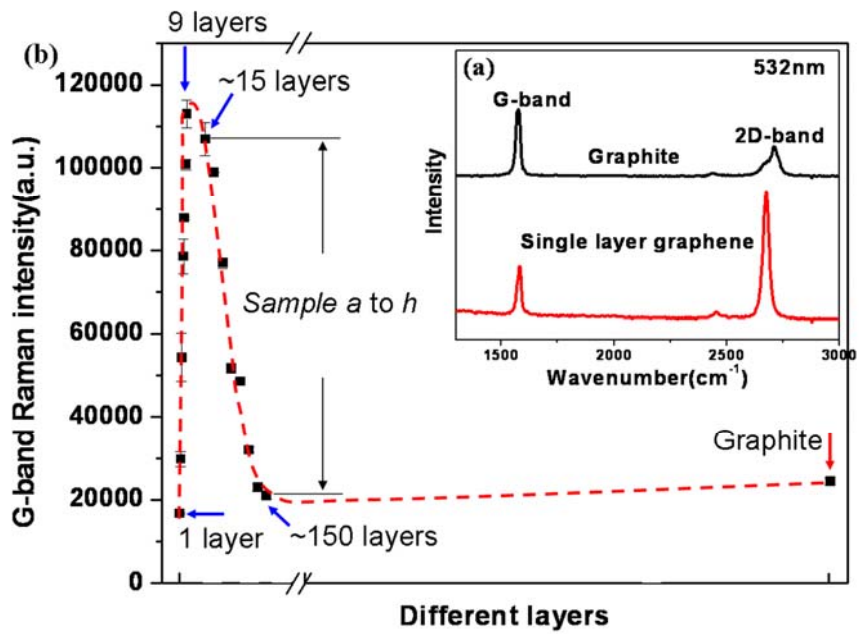

Figure 1. (a) The typical Raman spectra of graphene and graphite. (b) The G-band Raman intensity of graphene sheets as a function of number of layers. The red dashed curve is a guide to eye.

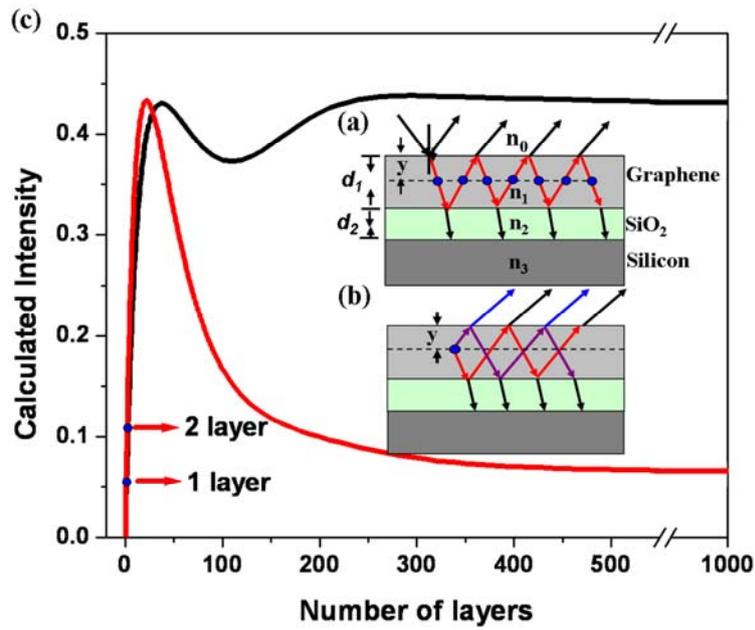

Figure 2. (a) Schematic laser reflection and transmission in certain depth *y* in graphene sheets deposited on $SiO_2$/Si substrate. (b) Multi-reflection of the scattering Raman light (from depth *y*) at the interface graphene/air and graphene/ ($SiO_2$ on Si). The detected Raman signal is a summation of infinite transmitted light at the interface of graphene/air. (c).The calculation results of Raman intensity of G band as a function of number of layers with (red) and without (black) considering the multi-reflection of Raman scattering light in graphene.

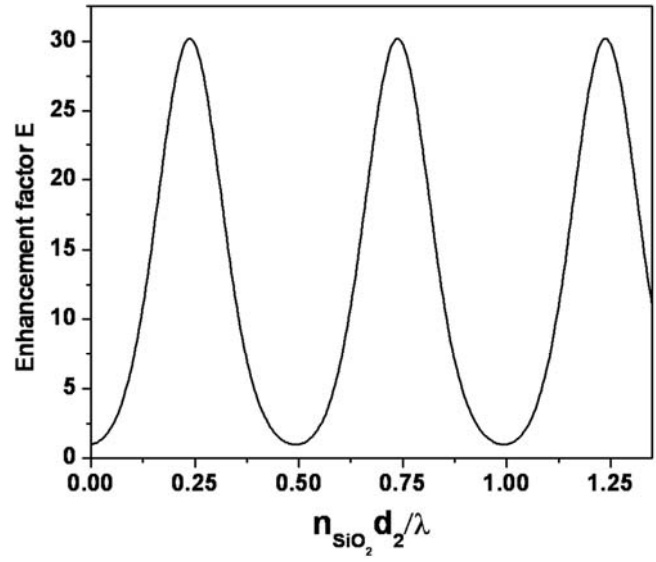

Figure 3. Calculation result of enhancement factor E as a function of $\dfrac{n_{SiO_2} \cdot d_2}{\lambda}$